\pdfoutput=1
\documentclass{article}

\usepackage[final]{neurips_2023}

\usepackage[utf8]{inputenc} 
\usepackage[T1]{fontenc}    
\usepackage[hidelinks]{hyperref} 
\usepackage{url}            
\usepackage{booktabs}       
\usepackage{amsfonts}       
\usepackage{nicefrac}       
\usepackage{microtype}      
\usepackage{xcolor}         
\usepackage[hang,flushmargin]{footmisc}
\usepackage{enumitem}
\usepackage{xspace}
\usepackage{arydshln}
\usepackage{multirow}
\usepackage{graphicx}

\usepackage{etoolbox}
\makeatletter
\patchcmd{\@verbatim}
  {\verbatim@font}
  {\verbatim@font\small}
  {}{}
\makeatother

\newcommand{\ignore}[1]{}

\title{Vector Search with OpenAI Embeddings:\\ Lucene Is All You Need}

\author{Jimmy Lin,$^1$ Ronak Pradeep,$^1$ Tommaso Teofili,$^2$ Jasper Xian$^1$ \\[1ex]
$^1$ David R. Cheriton School of Computer Science, University of Waterloo\\
$^2$ Department of Engineering, Roma Tre University
}

\begin{document}

\maketitle

\begin{abstract}
We provide a reproducible, end-to-end demonstration of vector search with OpenAI embeddings using Lucene on the popular MS MARCO passage ranking test collection.
The main goal of our work is to challenge the prevailing narrative that a dedicated vector store is necessary to take advantage of recent advances in deep neural networks as applied to search.
Quite the contrary, we show that hierarchical navigable small-world network (HNSW) indexes in Lucene are adequate to provide vector search capabilities in a standard bi-encoder architecture.
This suggests that, from a simple cost--benefit analysis, there does not appear to be a compelling reason to introduce a dedicated vector store into a modern ``AI stack'' for search, since such applications have already received substantial investments in existing, widely deployed infrastructure.
\end{abstract}


\section{Introduction}

Recent advances in the application of deep neural networks to search have focused on representation learning in the context of the so-called bi-encoder architecture, where content (queries, passages, and even images and other multimedia content) is represented by dense vectors (so-called ``embeddings'').
Dense retrieval models using this architecture form the foundation of retrieval augmentation in large language models (LLMs), a popular and productive approach to improving LLM capabilities in the broader context of generative AI~\citep{Mialon:2302.07842:2023,asai-etal-2023-retrieval}.

The dominant narrative today is that since dense retrieval requires the management of a potentially large number of dense vectors, enterprises require a dedicated ``vector store'' or ``vector database'' as part of their ``AI stack''.
There is a cottage industry of startups that are pitching vector stores as novel, must-have components in a modern enterprise architecture; examples include Pinecone, Weaviate, Chroma, Milvus, Qdrant, just to name a few.
Some have even argued that these vector databases will replace the venerable relational database.\footnote{\scriptsize \url{https://twitter.com/andy_pavlo/status/1659740200266870787}}

The goal of this paper is to provide a counterpoint to this narrative.
Our arguments center around a simple cost--benefit analysis:\ since search is a brownfield application, many organizations have already made substantial investments in these capabilities.
Today, production infrastructure is dominated by the broad ecosystem centered around the open-source Lucene search library, most notably driven by platforms such as Elasticsearch, OpenSearch, and Solr.
While the Lucene ecosystem has admittedly been slow to adapt to recent trends in representation learning, there are strong signals that serious investments are being made in this space.
Thus, we see no compelling reason why separate, dedicated vector stores are necessary in a modern enterprise.
In short, the benefits do not appear to justify the cost of additional architectural complexity.

It is important to separate the need for {\it capabilities} from the need for distinct {\it software components}.
While hierarchical navigable small-world network (HNSW) indexes~\citep{HNSW} represent the state of the art today in approximate nearest neighbor search---the most important operation for vector search using embeddings---it is not clear that providing operations around HNSW indexes requires a separate and distinct vector store.
Indeed, the most recent major release of Lucene (version 9, from December 2021) includes HNSW indexing and vector search, and these capabilities have steadily improved over time.
The open-source nature of the Lucene ecosystem means that advances in the core library itself will be rapidly adopted and integrated into other software platforms within the broader ecosystem.

The growing popularity of so-called embedding APIs~\citep{Kamalloo_etal_ACL2023} further strengthens our arguments.
These APIs encapsulate perhaps the most complex and resource-intensive aspect of vector search---the generation of dense vectors from pieces of content.
Embedding APIs hide model training, deployment, and inference behind the well-known benefits of service-based computing, much to the delight of practitioners.
To support our arguments, we demonstrate vector search with OpenAI embeddings~\citep{Neelakantan:2201.10005:2022} using the popular MS MARCO passage ranking test collection~\citep{msmarco}.
Specifically, we have encoded the entire corpus and indexed the embedding vectors using Lucene.
Evaluation on the MS MARCO development set queries and queries from the TREC Deep Learning Tracks~\citep{Craswell_etal_TREC2019_DL_overview,Craswell_etal_TREC2020_DL_overview} show that OpenAI embeddings are able to achieve a respectable level of effectiveness.
And as \cite{Devins_etal_WSDM2022} have shown, anything doable in Lucene is relatively straightforward to replicate in Elasticsearch (and any other platform built on Lucene).
Thus, we expect the ideas behind our demonstration to become pervasive in the near future.

We make available everything needed to reproduce the experiments described in this paper, starting with the actual OpenAI embeddings, which we make freely downloadable.\footnote{\scriptsize \url{https://github.com/castorini/anserini/blob/master/docs/experiments-msmarco-passage-openai-ada2.md}}
At a high-level, our demonstration shows how easy it is to take advantage of state-of-the-art AI techniques today without any AI-specific implementations per se:\ embeddings can be computed with simple API calls, and indexing and searching dense vectors is conceptually identical to indexing and searching text with bag-of-words models that have been available for decades.

\section{From Architecture to Implementation}

\begin{figure*}[t]
\begin{center}
\centerline{\includegraphics[width=0.6\textwidth]{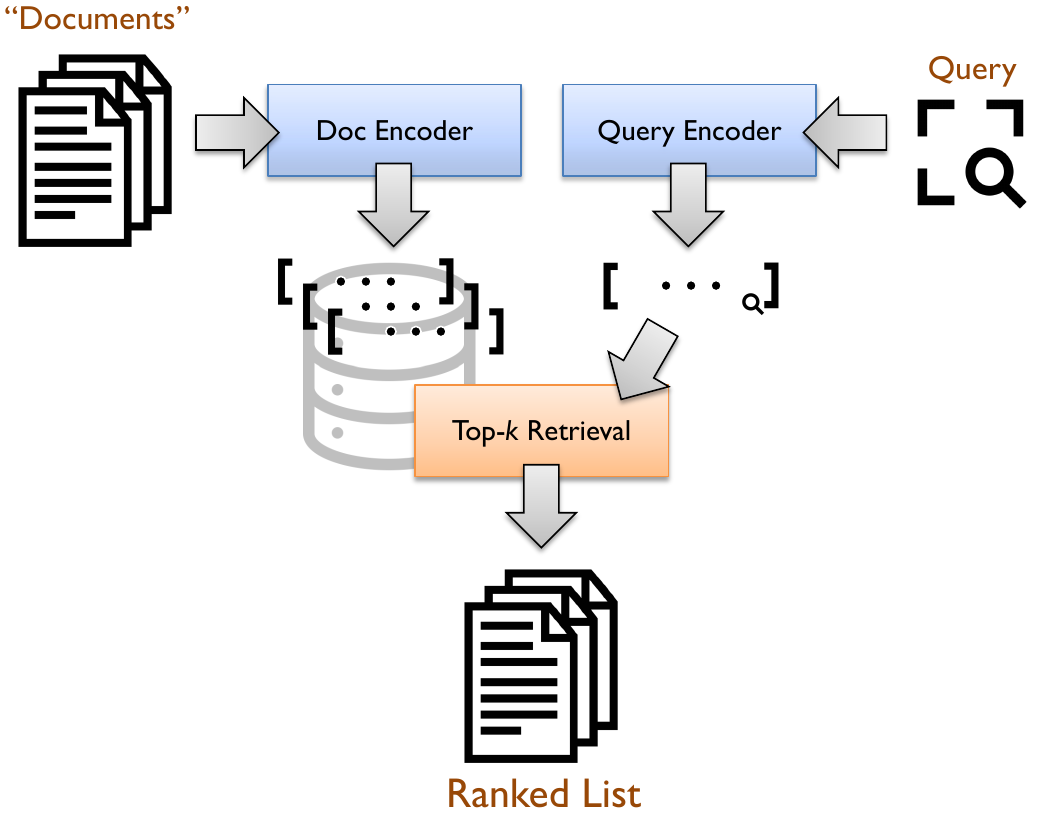}}
\caption{A standard bi-encoder architecture, where encoders generate dense vector representations (embeddings) from queries and documents (passages). Retrieval is framed as $k$-nearest neighbor search in vector space.} 
\label{bi-encoder}
\end{center}
\end{figure*}

The central idea behind the bi-encoder architecture (see Figure~\ref{bi-encoder}) is to encode queries and passages into dense vectors---commonly referred to as ``embeddings''---such that relevant query--passage pairs receive high scores, computed as the dot product of their embeddings.
In this manner, search can be reformulated as a nearest neighbor search problem in vector space:\ given the query embedding, the system's task is to rapidly retrieve the top-$k$ passage embeddings with the largest dot products~\citep{Lin_arXiv2021_repir}.
Typically, ``encoders'' for generating the vector representations are implemented using transformers, which are usually fine-tuned in a supervised manner using a large dataset of relevant query--passage pairs~\citep{dpr,Xiong_etal_ICLR2021}.

This formulation of search, in terms of comparisons between dense vectors, differs from ``traditional'' bag-of-words sparse representations that rely on inverted indexes for low-latency query evaluation.
Instead, nearest neighbor search in vector space requires entirely different techniques:\ indexes based on hierarchical navigable small-world networks (HNSW)~\citep{HNSW} are commonly acknowledged as representing the state of the art.
The Faiss library~\citep{faiss} provides a popular implementation of HNSW indexes that is broadly adopted today and serves as a standard baseline.
Despite conceptual similarities~\citep{Lin_arXiv2021_repir}, it is clear that top-$k$ retrieval on sparse vectors and dense vectors require quite different and distinct ``software stacks''.
Since hybrid approaches that combine both dense and sparse representations have been shown to be more effective than either alone~\citep{Ma_etal_ECIR2022,Lin_Lin_TOIS2023}, many modern systems combine separate retrieval components to achieve hybrid retrieval.
For example, the Pyserini IR toolkit~\citep{Lin_etal_SIGIR2021_Pyserini} integrates Lucene and Faiss for sparse and dense retrieval, respectively.

Recognizing the need for managing both sparse and dense retrieval models, the dominant narrative today is that the modern enterprise ``AI stack'' requires a dedicated vector store or vector database, alongside existing fixtures such as relational databases, NoSQL stores, event stores, etc.
A vector store would handle, for example, standard CRUD (create, read, update, delete) operations as well as nearest neighbor search.
Many startups today are built on this premise; examples include Pinecone, Weaviate, Chroma, Milvus, Qdrant, just to name a few.
This is the narrative that our work challenges.

Modern enterprise architectures are already exceedingly complex, and the addition of another software component (i.e., a distinct vector store) requires carefully weighing costs as well as benefits.
The cost is obvious:\ increased complexity, not only from the introduction of a new component, but also from interactions with existing components.
What about the benefits?
While vector stores no doubt introduce new capabilities, the critical question is whether these capabilities can be provided via alternative means.


Search is a brownfield application.
Wikipedia defines this as ``a term commonly used in the information technology industry to describe problem spaces needing the development and deployment of new software systems in the immediate presence of existing (legacy) software applications/systems.''
Additionally, ``this implies that any new software architecture must take into account and coexist with live software already in situ.''
Specifically, many organizations have already made substantial investments in search within the Lucene ecosystem.
While most organizations do not directly use the open-source Lucene search library in production, the search application landscape is dominated by platforms that are built on top of Lucene such as Elasticsearch, OpenSearch, and Solr.
For example, Elastic, the publicly traded company behind Elasticsearch, reports approximately 20,000 subscriptions to its cloud service as of Q4 FY2023.\footnote{\scriptsize \url{https://ir.elastic.co/news-events/press-releases/press-releases-details/2023/Elastic-Reports-Fourth-Quarter-and-Fiscal-2023-Financial-Results/default.aspx}}
Similarly, in the category of search engines, Lucene dominates DB-Engines Ranking, a site that tracks the popularity of various database management systems.\footnote{\scriptsize \url{https://db-engines.com/en/ranking/search+engine}}
There's a paucity of concrete usage data, but it would not be an exaggeration to say that Lucene has an immense install base.


The most recent major release of Lucene (version 9), dating back to December 2021, includes HNSW indexing and search capabilities, which have steadily improved over the past couple of years.
This means that differences in capabilities between Lucene and dedicated vector stores are primarily in terms of performance, not the availability of must-have features.
Thus, from a simple cost--benefit calculus, it is not clear that vector search requires introducing a dedicated vector store into an already complex enterprise ``AI stack''.
Our thesis:\ Lucene is all you need.

We empirically demonstrate our claims on the MS MARCO passage ranking test collection, a standard benchmark dataset used by researchers today.
We have encoded the entire corpus using OpenAI's \texttt{ada2} embedding endpoint, and then indexed the dense vectors with Lucene.
Experimental results show that this combination achieves effectiveness comparable to the state of the art on the development queries as well as queries from the TREC 2019 and 2020 Deep Learning Tracks.

Our experiments are conducted with Anserini~\citep{Yang_etal_JDIQ2018}, a Lucene-based IR toolkit that aims to support reproducible information retrieval research.
By building on Lucene, Anserini aims to bridge the gap between academic information retrieval research and the practice of building real-world search applications.
\cite{Devins_etal_WSDM2022} showed that capabilities implemented by researchers in Anserini using Lucene can be straightforwardly translated into Elasticsearch (or any other platform in the Lucene ecosystem), thus simplifying the path from prototypes to production deployments.

Our demonstration further shows the ease with which state-of-the-art vector search can be implemented by simply ``plugging together'' readily available components.
In the context of the bi-encoder architecture, \citet{Lin_arXiv2021_repir} identified the logical scoring model and the physical retrieval model as distinct conceptual components.
In our experiments, the logical scoring model maps to the OpenAI embedding API---whose operations are no different from any other API endpoint.
What Lin calls the physical retrieval model focuses on the top-$k$ retrieval capability, which is handled by Lucene.
In Anserini, vector indexing and search is exposed in a manner that is analogous to indexing and retrieval using bag-of-words models such as BM25.
Thus, the implementation of the state of the art in vector search using generative AI does not require any AI-specific implementations, which increases the accessibility of these technologies to a wider audience.

\section{Experiments}

Experiments in this paper are relatively straightforward.
We focused on the MS MARCO passage ranking test collection~\citep{msmarco}, which is built on a corpus comprising approximately 8.8 million passages extracted from the web.
Note that since the embedding vectors are generated by OpenAI's API endpoint, no model training was performed.
For evaluation, we used the standard development queries as well as queries from the TREC 2019 and TREC 2020 Deep Learning Tracks.

In our experimental setup, we utilized the OpenAI \texttt{ada2} model~\citep{Neelakantan:2201.10005:2022} for generating both query and passage embeddings.
This model is characterized by an input limit of 8191 tokens and an output embedding size of 1536 dimensions.
However, to maintain consistency with the existing literature~\citep{Pradeep_etal_arXiv2021_EMD,Ma_etal_SIGIR2022}, we truncated all passages in the corpus to 512 tokens.
It is unknown whether OpenAI leveraged the MS MARCO passage corpus during model development, but in general, accounting for data leakage is extremely challenging for large models, especially those from OpenAI that lack transparency.

Using \texttt{tiktoken}, OpenAI's official tokenizer, we computed the average token count per passage in our corpus to be 75.2, resulting in a total of approximately 660 million tokens.
In order to generate the embeddings efficiently, we queried the API in parallel while respecting the rate limit of 3500 calls per minute.
We had to incorporate logic for error handling in our code, given the high-volume nature of our API calls.
Ultimately, we were able to encode both the corpus and the queries, the latter of which are negligible in comparison, in a span of two days.

As previously mentioned, all our retrieval experiments were conducted with the Anserini IR toolkit~\citep{Yang_etal_JDIQ2018}.
The primary advantage of Anserini is that it provides direct access to underlying Lucene features in a ``researcher-friendly'' manner that better comports with modern evaluation workflows.
Our experiments were based on Lucene 9.5.0, but indexing was a bit tricky because the HNSW implementation in Lucene restricts vectors to 1024 dimensions, which was not sufficient for OpenAI's 1536-dimensional embeddings.\footnote{\scriptsize \url{https://github.com/apache/lucene/issues/11507}}
Although the resolution of this issue, which is to make vector dimensions configurable on a per codec basis, has been merged to the Lucene source trunk,\footnote{\scriptsize \url{https://github.com/apache/lucene/pull/12436}} this feature has not been folded into a Lucene release (yet) as of early August 2023.
Thus, there is no public release of Lucene that can directly index OpenAI's \texttt{ada2} embedding vectors.
Fortunately, we were able to hack around this limitation in an incredibly janky way.\footnote{The sketch of the solution is as follows:\ We copy relevant source files from the Lucene source trunk directly into our source tree and patch the vector size settings directly. When we build our fatjar, the class files of our ``local versions'' take precedence, and hence override the vector size limitations.}

Experimental results are shown in Table~\ref{results}, where we report effectiveness in terms of standard metrics:\ reciprocal rank at 10 (RR@10), average precision (AP), nDCG at a rank cutoff of 10 (nDCG@10), and recall at a rank cutoff of 1000 (R@1k).
The effectiveness of the \texttt{ada2} embeddings is shown in the last row of the table.
Note that due to the non-deterministic nature of HNSW indexing, effectiveness figures may vary slightly from run to run.

For comparison, we present results from a few select points of reference, classified according to the taxonomy proposed by~\cite{Lin_arXiv2021_repir};
OpenAI's embedding models belong in the class of learned dense representations.
Notable omissions in the results table include the following:\ the original OpenAI paper that describes the embedding model \citep{Neelakantan:2201.10005:2022} does not report comparable results; neither does~\cite{Izacard:2112.09118:2021} for Contriever, another popular learned dense representation model.
Recently, \cite{Kamalloo_etal_ACL2023} also evaluated OpenAI's \texttt{ada2} embeddings, but they did not examine any of the test collections we do here.
Looking at the results table, our main point is that we can achieve effectiveness comparable to the state of the art using a production-grade, completely off-the-shelf embedding API coupled with Lucene for indexing and retrieval.

\begin{table}[t]
\centering
\scalebox{0.87}{
\setlength{\tabcolsep}{3pt}
\begin{tabular}{lcccccccc}
\toprule
& \multicolumn{2}{c}{{\bf dev}} & \multicolumn{3}{c}{{\bf DL19}} & \multicolumn{3}{c}{{\bf DL20}}\\
& RR@10 & R@1k & AP & nDCG@10 & R@1k & AP & nDCG@10 & R@1k\\
\cmidrule(lr){2-3} \cmidrule(lr){4-6} \cmidrule(lr){7-9}
\multicolumn{4}{l}{{\bf Unsupervised Sparse Representations}} \\
BM25~\citep{Ma_etal_SIGIR2022}$^*$     & 0.184 & 0.853 & 0.301 & 0.506 & 0.750 & 0.286 & 0.480 & 0.786 \\
BM25+RM3~\citep{Ma_etal_SIGIR2022}$^*$ & 0.157 & 0.861 & 0.342 & 0.522 & 0.814 & 0.301 & 0.490 & 0.824 \\[1ex]
\multicolumn{4}{l}{{\bf Learned Sparse Representations}} \\
uniCOIL~\citep{Ma_etal_SIGIR2022}$^*$ & 0.352 & 0.958 & 0.461 & 0.702 & 0.829 & 0.443 & 0.675 & 0.843 \\
SPLADE++ ED~\citep{splade}$^*$ & 0.383 & 0.983 & 0.505 & 0.731 & 0.873 & 0.500 &	0.720 & 0.900 \\[1ex]
\multicolumn{4}{l}{{\bf Learned Dense Representations}} \\
TAS-B~\citep{Hofstatter_etal_SIGIR2021} & 0.340 & 0.975 & - & 0.712 & 0.845 & - & 0.693 & 0.865 \\
TCT-ColBERTv2~\citep{Lin_etal_2021_RepL4NLP}$^*$ & 0.358 & 0.970 & 0.447 & 0.720 & 
0.826 & 0.475	& 0.688 & 0.843 \\
ColBERT-v2~\citep{santhanam-etal-2022-colbertv2} & 0.397 & 0.984 & - & - & - & - & - & - \\
Aggretriever~\citep{Lin_etal_TACL2023}$^*$ & 0.362 & 0.974 & 0.435 & 0.684 & 0.808 & 0.471 & 0.697 & 0.856 \\
\midrule
OpenAI \texttt{ada2} & 0.343 & 0.984 & 0.479 & 0.704 & 0.863 & 0.477 & 0.676 & 0.871 \\
\bottomrule
\end{tabular}
}
\vspace{0.2cm}
\caption{Effectiveness of OpenAI \texttt{ada2} embeddings on the MS MARCO development set queries (dev) and queries from the TREC 2019/2020 Deep Learning Tracks (DL19/DL20), compared to a selection of other models. $^*$ indicates results from Pyserini's two-click reproductions~\citep{Lin_arXiv2022} available at {\small \url{https://castorini.github.io/pyserini/2cr/msmarco-v1-passage.html}}, which may differ slightly from the original papers. All other results are copied from their original papers.}
\label{results}
\end{table}

To complete our experimental results, we provide performance figures on a server with two Intel Xeon Platinum 8160 processors (33M Cache, 2.10 GHz, 24 cores each) with 1 TB RAM, running Ubuntu 18.04 with ZFS.
This particular processor was launched in Q3 of 2017 and is no longer commercially available; we can characterize this server as ``high end'', but dated.
Indexing took around three hours with 16 threads, with the parameters \texttt{M} set to 16 and \texttt{efC} set to 100, without final segment optimization.
Using 32-bit floats, the raw 1536-dimensional vectors should consume 54 GB on disk, but for convenience we used an inefficient JSON text-based representation.
Therefore, our collection of vectors takes up 109 GB as compressed text files (using gzip).
For vector search, using 16 threads, we were able to achieve 9.8 queries per second (QPS), fetching 1000 hits per query with the \texttt{efSearch} parameter set to 1000.
These results were obtained on the MS MARCO development queries, averaged over four separate trials after a warmup run.

\section{Discussion}

Our demonstration shows that it is possible today to build a vector search prototype using OpenAI embeddings directly with Lucene.
Nevertheless, there are a number of issues worth discussing, which we cover below.

\paragraph{Jank.}
We concede that getting our demonstration to work required a bit of janky implementation tricks.
Even though all the required features have been merged to Lucene's source trunk, no official release has been cut that incorporates all the patches (at least at the time we performed our experiments in early August, 2023).
Quite simply, the complete feature set necessary for production deployment is not, as they say, ready for prime time.
However, to use another clich\'{e}, this is a small matter of programming (SMOP).
We see no major roadblocks in the near future:\ the next official release of Lucene will incorporate the necessary features, and after that, all downstream consumers will begin to incorporate the capabilities that we demonstrate here.

Nevertheless, Lucene has been a relative laggard in dense retrieval.
Despite this, we believe that recent developments point to substantial and sustained investments in the Lucene ecosystem moving forward.
For example, in its Q4 FY 2023 report, Elastic announced the Elasticsearch Relevance Engine, ``powered by built-in vector search and transformer models, designed specifically to bring the power of AI innovation to proprietary enterprise data.''
A recent blog post\footnote{\scriptsize \url{https://aws.amazon.com/blogs/big-data/amazon-opensearch-services-vector-database-capabilities-explained/}} from Amazon Web Services explained vector database capabilities in OpenSearch, providing many details and reference architectures.
These are just two examples of commitments that help bolster the case for Lucene that we have articulated here.
Overall, we are optimistic about the future of the ecosystem.

\paragraph{Performance.}
Lucene still lags alternatives in terms of indexing speed, query latency and throughput, and related metrics.
For example, \cite{Ma_etal_CIKM2023} recently benchmarked Lucene 9.5.0 against Faiss~\citep{faiss}.
Experiments suggest that Lucene achieves only around half the query throughput of Faiss under comparable settings, but appears to scale better when using multiple threads.
Although these results only capture a snapshot in time, it would be fair to characterize Lucene as unequivocally slower.
However, Faiss is relatively mature and hence its headroom for performance improvements is rather limited.
In contrast, we see many more opportunities for gains in Lucene.
Coupled with signs of strong commitment (discussed above), we believe that the performance gap between Lucene and dedicated vector stores will decrease over time.

\paragraph{Alternatives.}
We acknowledge a number of competing alternatives that deserve consideration.
Note that the core argument we forward is about cost--benefit tradeoffs:
In our view, it is not clear that the benefits offered by a dedicated vector store outweigh the increased architectural complexity of introducing a new software component within an enterprise.
From this perspective, we can identify two potentially appealing alternatives:

\begin{itemize}[leftmargin=0.5cm]

\item {\it Fully managed services.}
One simple way to reduce architectural complexity is to make it someone else's problem.
Vespa\footnote{\scriptsize \url{https://vespa.ai/}} is perhaps the best example of this solution, providing both dense retrieval and sparse retrieval capabilities in a fully managed environment, eliminating the need for users to explicitly worry about implementation details involving inverted indexes, HNSW indexes, etc.
Vepsa provides a query language that supports a combination of vector search, full-text search, as well as search over structured data.
Our main question here concerns traction and adoption:\ as a brownfield application, we're not convinced that enterprises will make the (single, large) leap from an existing solution to a fully managed service.

\item {\it Vector search capabilities in relational databases.}
In the same way that vector search grows naturally out of an already deployed and mature text search platform (e.g., Elasticsearch), we can see similar arguments being made from the perspective of relational databases.
Despite numerous attempts (spanning decades) at toppling its lofty perch~\citep{goesaround,Pavlo_etal_SIGMOD2009}, relational databases remain a permanent fixture in enterprise ``data stacks''.
This means that by building vector search capabilities into relational databases, enterprises gain entr\'{e}e into the world of dense retrieval (essentially) for free.
A great example of this approach is pgvector,\footnote{\scriptsize \url{https://github.com/pgvector/pgvector}} which provides open-source vector similarity search for Postgres.
We find the case compelling:\ if your enterprise is already running Postgres, pgvector adds vector search capabilities with minimal additional complexity.
It's basically a free lunch.

\end{itemize}

\section{Conclusions}

There is no doubt that manipulation of dense vectors forms an important component of search today.
The central debate we tackle is how these capabilities should be implemented and deployed in production systems.
The dominant narrative is that you need a new, distinct addition to your enterprise ``AI stack''---a vector store.
The alternative we propose is to say:\ If you've built search applications already, chances are you're already invested in the Lucene ecosystem.
In this case, Lucene is all you need.
Of course, time will tell who's right.

\section*{Acknowledgements}

This research was supported in part by the Natural Sciences and Engineering Research Council (NSERC) of Canada.
We'd like to thank Josh McGrath and the team at Distyl for providing support to access OpenAI APIs.

\bibliographystyle{ACM-Reference-Format}
\bibliography{hnsw}

\end{document}